\documentstyle[psfig]{proceeding}
\newcommand{\ltap}{\mathrel{\hbox{\rlap{\lower.55ex \hbox {$\sim$}}
                   \kern-.3em \raise.4ex \hbox{$<$}}}}
\newcommand{\gtap}{\mathrel{\hbox{\rlap{\lower.55ex \hbox {$\sim$}}
                   \kern-.3em \raise.4ex \hbox{$>$}}}}

\begin{document}
\title{X-ray sources in old stellar clusters: the contribution of ROSAT}
\author{Frank Verbunt}  
\institute{Astronomical Institute,
 Postbox 80\,000, 3508~TA~Utrecht, the Netherlands (verbunt@phys.uu.nl)}
\maketitle

\begin{abstract}
Two new bright X-ray sources in globular clusters, and many less
luminous ones in globular clusters and in old open clusters, have been
discovered with ROSAT.
Accurate positions obtained with ROSAT help identification with
optical objects, which however is still very incomplete in globular
clusters. One dim globular cluster source has been identified with 
a recycled radio pulsar; several others may be cataclysmic variables.
The four brightest X-ray sources identified in the old open cluster M$\,$67
are puzzling, as we do not understand why they emit X-rays.
In comparison with the two old open clusters studied so far, globular
clusters are remarkably underluminous in X-rays.
\end{abstract}

\section*{0. First}

I am grateful to the organizers of this meeting
for the opportunity to contribute to this
celebration of the 65$^{\rm th}$ birthday of Joachim Tr\"umper.
During my years (1985-1989) at the Max Planck Institut f\"ur
extraterrestrische Physik, I have always felt very much at home,
as if an `honorary German',
and I thank Joachim and my other colleagues for this. 
I was impressed with the width and depth of Joachim Tr\"umper's
interests, and thoroughly enjoyed my many discussions with him.

The deadline for the first Announcement of Opportunity for
observing with ROSAT was on a Sunday evening, at midnight
(in checking the date, I find it was on my birthday!). Joachim Tr\"umper
was in his office that night, to sign proposals until just before the
deadline, or almost, as 11 copies needed to be made for submission.
Some of the proposals he signed that night contributed to the results
discussed in this review.

\section{Introduction}

Two types of stellar clusters are commonly discriminated.
{\em Globular clusters} are distributed spherically around the
Galactic Center, have metal abundances $\ltap 0.1$ of the solar
abundances, contain each up to $10^6-10^7$ stars, and are 
very old (in danger in fact of being older than the Universe...),
$11.5\pm1.5$\,Gyr (Chaboyer et al.\ 1998).
{\em Open} or {\em galactic clusters} are located in the disk
of the Milky Way, have metallicities comparable to solar, contain
each some $10^3-10^4$ stars, and in age range from very young ($\sim$\,Myr)
to fairly old $\ltap6$\,Gyr).
Open clusters have been found only relatively nearby, and it is quite
possible that open clusters exist with star numbers and ages closer
to those of globular clusters.

The study of clusters with different ages has taught us much about the 
evolution of single stars; with large numbers of binaries now being detected in
clusters we may hope to learn much also about the evolution of binaries.
There is an interesting interplay between the evolution
of a cluster as a whole and the evolution of the individual stars in it
(as reviewed by Hut et al.\ 1992).
For example, sudden mass loss by many of the stars (e.g.\ due to
supernovae in a very young cluster) can dissolve the cluster.
Close encounters between single stars and binaries can increase
the velocity dispersion of the stars in the cluster.
And finally, collisions between stars or exchange encounters (in which
a single star kicks a binary star out of its orbit and takes its place)
can lead to peculiar stars and binaries that could not arise from
isolated (binary) evolution.

The high stellar density in globular clusters implies that encounters
between stars in them are common; the lower density in open clusters 
imply that such encounters are less frequent there, although they may
have occurred more often in the past.
In this article I will illustrate how X-ray observations with ROSAT
contribute to the study of and comparison between binaries in 
globular clusters and in old open clusters.
X-ray emission identifies stars which are, or have been, interacting
with other stars.
Followup studies at optical and ultraviolet wavelengths may explain
the nature of the interaction.

In Sect.\ 2 I discuss the observations of globular clusters and the
detection of bright and of less luminous X-ray sources in them.
47$\,$Tuc and NGC$\,$6397 are discussed as examples of efforts
to optically identify dim X-ray sources in the cores of globular
clusters.
In Sect.~3 I discuss the discovery of X-ray sources in the old open
clusters M$\,$67 and NGC$\,$188, and of followup observational
and theoretical work.
Sect.\ 4 summarizes the conclusions.

\begin{figure}
\psfig{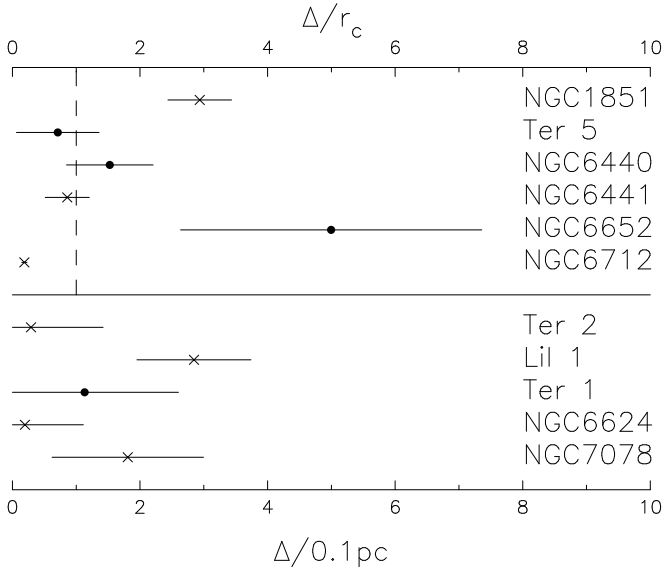}
\caption[]{Distances $\Delta$ to the cluster centers for the bright
sources in globular clusters. For ordinary clusters (above) the distance
is in units of the core radii $r_c$; for collapsed clusters (below) in units of
0.1$\,$pc. $\bullet$ indicates a position determination with ROSAT.
For NGC$\,$6440 it has been assumed that the dim source is the 
transient in quiescence.
Adapted from Johnston et al.\ (1995).}
\end{figure}
\nocite{jvh95}

\section{Globular clusters}

\subsection{Bright X-ray sources}

Since the discovery of the first bright ($L_x\gtap10^{36}\,$erg/s)
X-ray sources in globular clusters with UHURU, twelve such sources have been
found, five of which are transients (see the review by Hut et al.\ 1992). 
An X-ray source at these luminosities must be a neutron star or a black
hole accreting from a binary companion.
X-ray bursts due to thermonuclear fusion on the surface of a neutron
star have been detected in ten cluster sources, the last one with
SAX (in't Zand et al. 1998b).\footnote{After the meeting, a burst was
discovered of the X-ray source in NGC\,6640, making this the 11th 
system known to harbour a neutron star (in't Zand et al.\ 1998a).}
Binary periods have been found for four sources, the last one in 
NGC$\,$6712 (Homer et al.\ 1996).
Remarkably, two of these are extremely short, 11.4 min (source
in NGC\,6624) and 13.2 or 20.6 minutes (in NGC\,6712), 
indicating that the mass donor is a white dwarf with mass
$M_{wd}<0.1M_{\odot}$.
No such short-period X-ray binaries have been found in the Galactic Disk.
Repeated ROSAT observations show that the lightcurve of the 11.4 minute
binary is variable, complicating the effort to derive a period derivative
for this system (Van der Klis et al.\ 1993).

The globular cluster system contains only $\sim0.1$\%\ of the stars
of our Galaxy, but 10\%\ of the bright X-ray sources.
This indicates that bright X-ray sources are formed in globular clusters 
by stellar collisions or by exchange encounters between neutron stars and 
binaries.
\nocite{hmg+92}\nocite{zvh+98}\nocite{zhb+98}\nocite{hcn+96}

In agreement with this suggestion is the observation that the bright X-ray 
sources are located in the cores of the clusters.
ROSAT has increased the number of accurate source positions on which
this statement is based (Figure~1).

\begin{figure}
\psfig{figure=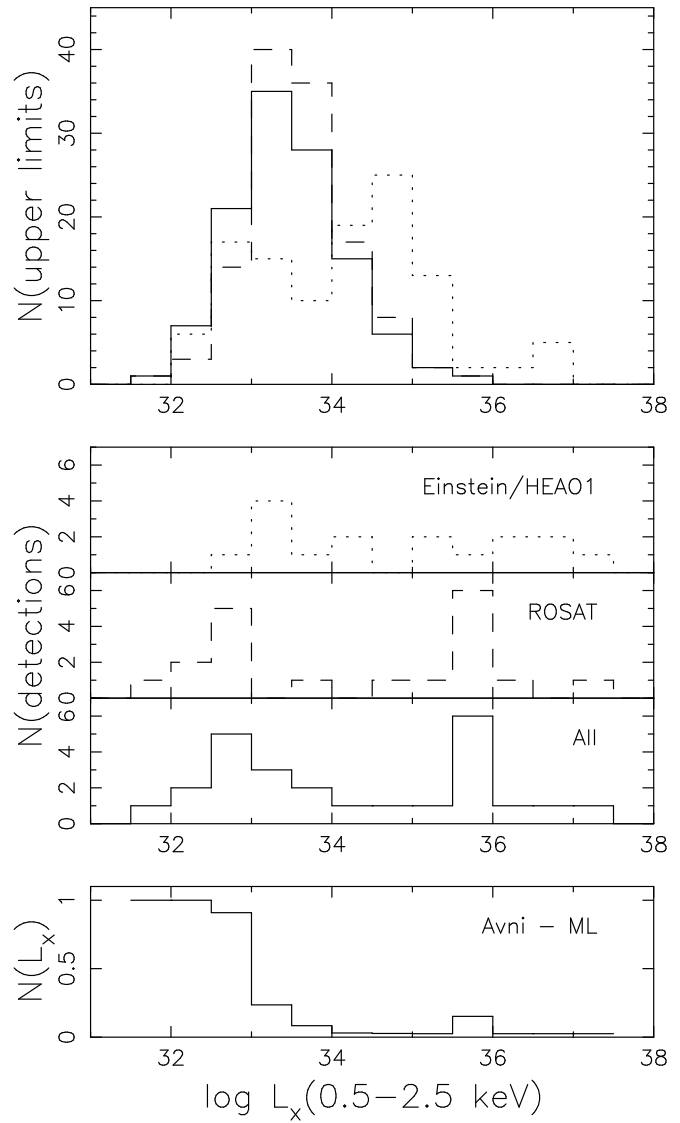,width=\columnwidth,clip=true}
\caption[]{Luminosity distributions of the cores of globular clusters,
showing detections (middle frame) and upper limits (upper frame),
as well as the derived most-likely intrinsic luminosity function (lower frame).
Luminosities refer to the 0.5-2.5$\,$keV band; pre-ROSAT data 
(from Einstein and HEAO1) are shown as dotted lines, ROSAT data as 
dashed lines and the combined data as solid lines.
From Verbunt et al.\ (1995).}
\end{figure}
\nocite{vbhj95}

\subsection{Dim X-ray sources}

\subsubsection{Census of dim sources}

The Einstein satellite discovered low-luminosity or dim 
($L_x\ltap10^{35}\,$erg/s) X-ray sources in the cores of nine
clusters, one of which (in NGC$\,$6440) may be the quiescent counterpart
of the bright transient in this cluster 
(Hertz \&\ Grindlay 1983).\footnote{After the meeting I analyzed a
long ROSAT HRI observation of NGC\,6440 obtained in 1993 and found that the dim
source in this cluster is in fact double; this will be discussed
in a paper on the 1998 outburst of the transient in NGC\,6440
(Verbunt et al.\ 1998).}\nocite{hg83}
Sources were also discovered outside the cores of $\omega\,$Cen
and NGC$\,$6656 but these have turned out to be fore- or background sources
(Cool et al.\ 1995a).\nocite{cgb+95}
ROSAT observations added substantially to this sample, bringing the total
to more than thirty (for a list, see Johnston \&\ Verbunt 1996), \nocite{jv96}
and has found multiple sources in the cores of NGC$\,$6397
(Cool et al.\ 1993), 47$\,$Tuc (Hasinger et al.\ 1994, Verbunt
 \&\ Hasinger 1998), $\omega\,$Cen (Johnston et al. 1994),
and NGC$\,$6652 (Bailyn 1995).
\nocite{cgkb93}\nocite{hjv94}\nocite{vh98}\nocite{jvh94}\nocite{bai95}

Combination of the detections with the upper limits (many of which
are from the Rosat All Sky Survey) gives the most likely X-ray
luminosity distribution of these sources, $dN(L_x)\propto {L_x}^{-1.5}dL_x$,
down to $\sim 10^{31.5}\,$erg/s (Figure 2).
This distribution implies that the total X-ray luminosity of each
cluster is dominated by a few bright sources rather than by large
numbers of undetected less luminous ones (Johnston \&\ Verbunt 1996).
This is confirmed by the observation of the unresolved X-ray flux
in spatially resolved cores, which is low or even undetectable.
Several sources have been shown to be variable, including one with
an extremely soft spectrum ($kT\sim40\,$eV) in M$\,$3 $=$ NGC$\,$5272
(Hertz et al.\ 1993).\nocite{hgb93}

The number $N$ of sources in  a cluster scales with core mass $M_c$ and density
of its core $\rho_c$ as $N\propto M_c{\rho_c}^{0.5}$ 
(Johnston \&\ Verbunt 1996).
This dependence of the number of sources is between the
proportionalities with mass $M_c$ and that expected for pure 
tidal capture $M_c\rho_c$, a result that was earlier found to hold for
radio pulsars in globular clusters (Johnston et al.\ 1992).\nocite{jkp92}
This may be the consequence of the destruction of binaries formed by
close encounters in subsequent encounters in the densest clusters; 
or because the sources are a mixture of binaries evolved from 
primordial binaries and binaries formed by close encounters.

\subsubsection{Nature of the dim sources}

The cores of globular clusters are known to harbour soft X-ray transients
in the low state (Sect.\ 2.1), cataclysmic variables, and millisecond radio 
pulsars; they probably also harbour chromospherically active close
binaries, RS~CVn systems.
All of these have been suggested as possible counterparts for the dim
X-ray sources.
One way to investigate these suggestions is by comparing the X-ray
luminosities of the dim sources in globular clusters with those
of individually identified
quiescent transients, cataclysmic variables, millisecond radio pulsars
and RS~CVn systems in the Galactic Disk (Figure~3).
It is seen that the relatively bright X-ray sources,
at luminosities $L_x\gtap10^{32}\,$erg/s) are matched in
luminosity only by the quiescent X-ray transients.
Millisecond radio pulsars, cataclysmic variables and RS~CVn systems
are found in the Galactic Disk at $L_x\ltap10^{32}\,$erg/s.
Indeed, at least part of the X-ray emission of the globular cluster
M$\,$28 is due to the millisecond pulsar in it (Danner et al.\ 1994, 1997).
\nocite{dkt94}\nocite{dksk97}

\begin{figure}
\psfig{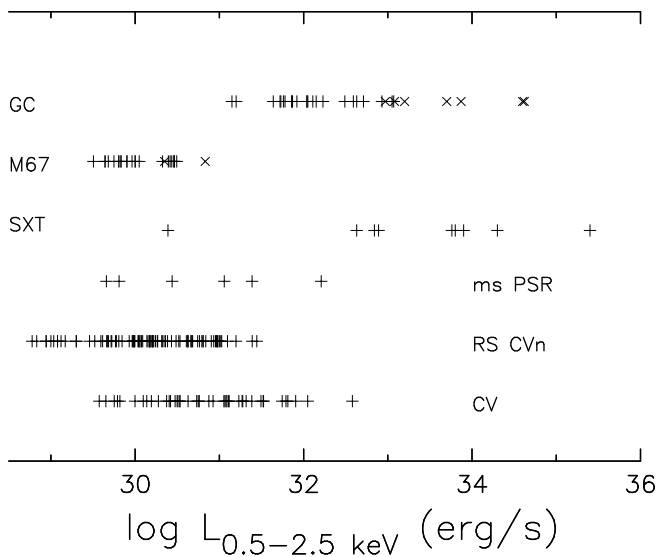}
\caption[]{X-ray luminosities of dim sources in globular clusters
($+$ spectrum known; $\times$ spectrum not known, conversion factor
counts to flux assumed) compared to those of sources in M$\,$67 ($+$)
and NGC$\,$188 ($\times$),  and of soft X-ray transients, millisecond 
radio pulsars, RS CVn systems and cataclysmic variables in the
galactic disk.
Adapted from Verbunt et al.\ (1997).}
\end{figure}
\nocite{vbrp97}

Dim X-ray sources have also been detected at positions compatible
with the known dwarf nova in M$\,$5 ($=$ NGC$\,$5904) and the known
old nova T$\,$Sco 1860 in M$\,$80  ($=$ NGC$\,$6093); if these are indeed the
counterparts the X-ray to optical flux ratio of the dwarf nova in M$\,$5
falls in the range of cataclysmic variables in the Galactic Disk, but
that of T$\,$Sco 1860 is rather higher (Figure~4).

\begin{figure}
\psfig{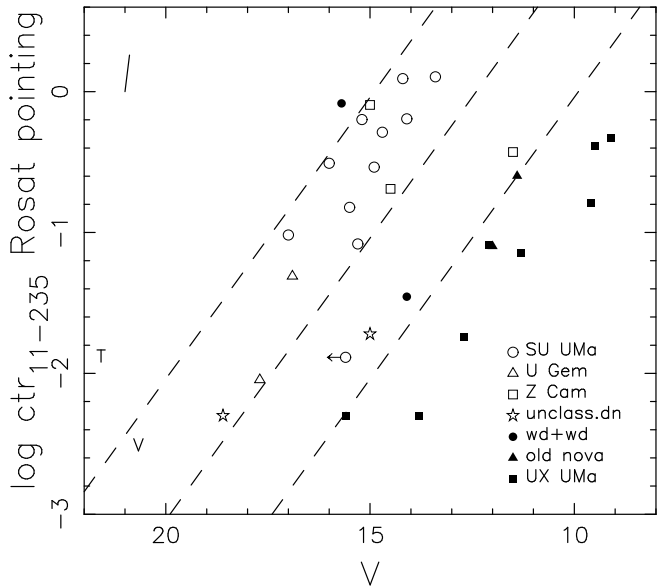}
\caption[]{The ROSAT PSPC countrate as a function of visual magnitude
for two known cataclysmic variables in globular clusters -- assuming
that these may be identified with the X-ray sources in these clusters --
and for various types of cataclysmic variables in the galactic disk. 
T is T$\,$Sco 1860 in M$\,$80, V is V101 in M$\,$5.
The dashed lines indicate constant ratios of X-ray flux to optical flux.
From Hakala et al.\ (1997).}
\end{figure}
\nocite{hcjv97}

Efforts to identify the multiple dim sources in 47$\,$Tuc and NGC$\,$6397
are described in the following subsections.

\begin{figure}
\psfig{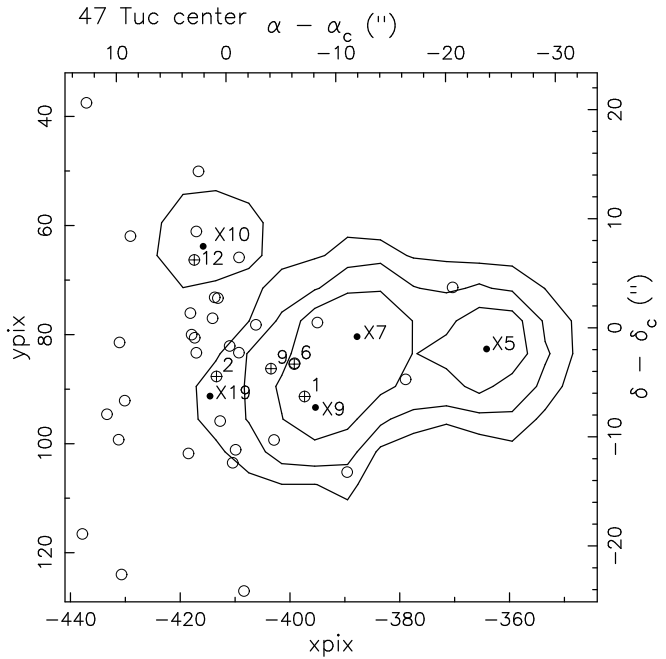}
\caption[]{X-ray contour levels of the core of 47~Tuc, with the
best locations of five X-ray sources ($\bullet$). Open circles indicate 
blue stragglers and variable stars listed by Geffert et al.\ (1997); these
are on an optical coordinate system, which may be shifted with respect to
the contours by up to 2$''$. Open circles inscribed with a plus sign are
discussed in the text.
The cataclysmic variables V$\,$1 and V$\,$2 (indicated with 1,2) are candidate
counterparts for X-ray sources X$\,$9 and X$\,$19; the eclipsing variable
12 from Edmonds et al.\ (1996) is a candidate counterpart for
X$\,$10. 6 and 9 indicate the variables AKO$\,$6,9 from Auri\`ere et 
al.\ (1989), which do not correspond to individually detected X-ray sources.
From Verbunt \&\ Hasinger (1998).}
\end{figure}
\nocite{gak97}\nocite{ako89}\nocite{egg+96}

\subsection{47$\,$Tuc}

The globular cluster 47$\,$Tuc  ($=$ NGC$\,$104) has been observed six times 
with the ROSAT HRI between 1992 and 1996.
Using four X-ray sources -- not related to the cluster -- detected in most
of these observations to align them precisely, Verbunt \&\ Hasinger (1998)
obtain an added 58,000$\,$s image of the core of 47$\,$Tuc.
Contour plots of this image are shown in Figure~5.
A multiple-source algorithm detects five significant sources in the core.

Two of the four sources outside the cluster which were used in the alignment
can be identified optically (with a G5 V star and with a galaxy,
Geffert et al.\ 1997), which
provides a tie of the X-ray coordinate frame to
the optical J$\,$2000 coordinate system with an estimated accuracy
of $\ltap 2''$. 
This reduces the number of possible optical counterparts to
the dim X-ray sources in the core, as compared to the number previously
allowed by the $5''$ positional accuracy.
In particular, the remarkable ultraviolet variable and binary AKO$\,$9
(Minniti et al.\ 1997)\nocite{mmp+97}
is not a counterpart to any of the individually detected X-ray sources.

Looking for counterparts among the blue stragglers, blue variables,
and eclipsing binaries, we find that the suggested cataclysmic variables
V$\,$1 and V$\,$2 are possible counterparts for X$\,$9 and X$\,$19
respectively, whereas an eclipsing binary, star 12 from Edmonds et al.\ (1996),
is a possible
counterpart to X$\,$10. It should be noted, however, that there is
a sizable probability that all coincidences are due to chance:
for the three X-ray sources in the central $20''\times20''$ region of the 
cluster, 22 counterparts are investigated, each acceptable as counterpart
in an area $4''\times4''$.

\begin{figure}
\psfig{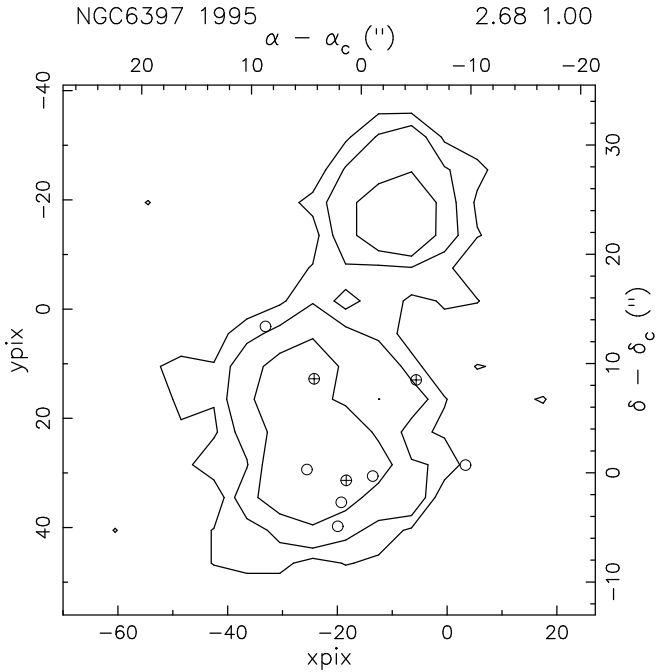}
\caption[]{X-ray contour levels of the core of NGC$\,$6397 as observed
with ROSAT HRI in 1995.
Circles indicate blue stars listed by Cool al.\ (1995b); those with 
H$\,\alpha$ emission are inscribed with $+$. 
}
\end{figure}

\subsection{NGC$\,$6397}

The globular cluster NGC$\,$6397 also contains multiple sources in the
center, as shown by the X-ray contours in Figure~6.  Cool et al.\
(1993) analyze the 1991-1992 data, and conclude that there are (at least)
three sources close to the core; another source is about 25$''$ to the
North of the cluster center.  Three H$\,\alpha$ emission stars, and
various blue stars, are suggested as possible counterparts by Cool 
et al.\ (1995b). Spectroscopy with the Hubble Space Telescope shows
that the H$\,\alpha$ emission stars have spectra like (magnetic) cataclysmic
variables (Grindlay et al.\ 1995).
\nocite{cgkb93}\nocite{gcc+95}\nocite{cgc+95}

In Figure~6 I show the optical positions of the proposed counterparts
on top of the X-ray contours, in a hitherto unpublished 1995 observation.
Within the likely error of $\sim 2''$ two H$\,\alpha$ candidates
are acceptable; the third (rightmost) one is less obvious.

\section{Old open clusters}

\subsection{M$\,$67}

Following the discovery of a cataclysmic variable of the AM~Her type in the
old open cluster M$\,$67 (Gilliland et al.\ 1991), ROSAT observations
of this cluster were obtained, which did indeed detect the variable,
and in addition detected some twenty other cluster members
(Belloni et al.\ 1993, 1998). \nocite{gbd+91}\nocite{bvs93}\nocite{bvm98}
The brightest X-ray sources in the ROSAT band are listed in Table~1.
They are all proper-motion and radial-velocity members of M$\,$67.
The location of identified X-ray sources in the colour magnitude 
diagram of M$\,$67 is shown in Figure~7.
The X-ray emission of several of these sources can be explained 
in terms of known source categories.
One very soft X-ray source, for example, is a hot white dwarf
(Pasquini et al.\ 1994, Fleming et al.\ 1997).\nocite{pba94}\nocite{flbb97}
Two sources are contact binaries whose X-ray emission is due to magnetic
activity.
Magnetic activity is also likely to explain the X-ray emission
of three binaries with circular orbits and binary periods $P_b\ltap10\,$d
(Belloni et al.\ 1998).

However, there are several sources whose X-ray emission is unexplained.
Remarkably, none of the four brightest sources in M$\,$67 is a 
predicted type of X-ray source.
All four are in locations in the  colour magnitude diagram away from
the isochrone of single M$\,$67 members, i.e.\ they are special stars
photometrically.

\begin{table}
\caption[o]{The eleven brightest X-ray sources in M$\,$67 (all with membership
probability $\geq86$\%), in order of X-ray luminosity. 
For each X-ray source counterpart the magnitude and colour are listed, the
X-ray luminosity in $0.1-2.4\,$keV band, orbital period,
orbital eccentricity, and the binary type. All X-ray luminosities
assume an X-ray spectrum typical for emission of a magnetically active
star. SB indicates a spectroscopic binary from the survey of Latham and
Mathieu (see Mathieu et al.\ 1990) without a determined orbit solution.  
Adapted from Belloni et al.\ (1998).
}
\begin{flushleft}
\begin{tabular}{rrrrrl}
$M_{\rm V}$            &
B-V          &
$L_{\rm x}$(erg/s)   &
$P_{\rm b}$(d) &
$e$            &
type      \\
\hline
11.52 & 0.88 &$7.8\times 10^{30}$&  42.83&0.00&giant$+$white dwarf\\
13.52 & 1.07 &$7.3\times 10^{30}$&  18.39&0.22&sub-subgiant\\
13.59 & 1.08 &$7.3\times 10^{30}$&2.82&0.03 &sub-subgiant \\
11.25 & 0.41 &$7.2\times 10^{30}$&     &     &blue straggler\\
12.47 & 0.66 &$6.8\times 10^{30}$&SB&     & RS CVn?\\
13.27 & 0.60 &$6.5\times 10^{30}$&   \\
14.38 & 0.73 &$6.2\times 10^{30}$&SB&     &RS CVn?\\
12.60 & 0.78 &$5.3\times 10^{30}$&  10.06&0.00&RSCVn\\
13.34 & 0.80 &$2.8\times 10^{30}$&   0.36&0.00&W Uma\\
15.03 & 0.81 &$2.5\times 10^{30}$&    &     &RS CVn??\\
13.98 & 0.61 &$2.5\times 10^{30}$&   2.66&0.00&RS CVn\\
\end{tabular}
\end{flushleft}
\end{table}
\nocite{mlg90}

\subsubsection{S$\,$1040: giant and white dwarf}

The brightest source is a binary that consists of a giant star and a
white dwarf, in a circular orbit around one another. It is star
1040 in Sanders' (1977) catalogue. \nocite{san77}\nocite{vp95}
Verbunt \&\ Phinney (1995) showed that the currently visible giant cannot
have circularized the orbit, if the standard circularization theory
is correct, and predicted that the unseen companion of the giant
would be a white dwarf, whose progenitor circularized the binary in
its giant stage.
This white dwarf was subsequently found in the ultraviolet; it is
undermassive, at $\sim 0.2M_{\odot}$, and has an effective temperature
$T_{\rm eff}=16160\,$K (Landsman et al.\ 1997).
\nocite{lab+97}

The white dwarf is too cool to be responsible for the X-ray emission.
The ultraviolet spectrum shows the Mg$\,$II$\,\lambda$2800 line in
emission (Landsman et al.), and a blue optical spectrum shows 
emission cores in the Ca$\,$II$\,$H and K lines (Van den Berg et al.\ 1998b),
both a sign of chromospheric activity on the giant star.\nocite{bvm98a}
This indicates that the X-ray emission is due to magnetic activity;
however, S\,1040 doesn't rotate fast, and it is not clear why it would
be magnetically active.

\begin{figure}
\psfig{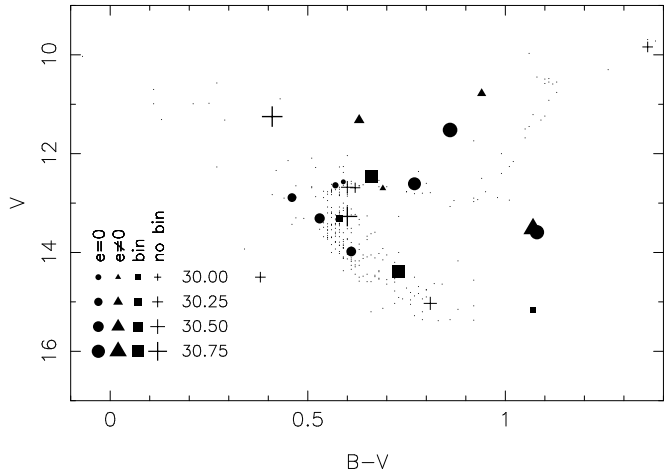}
\caption[]{Colour-magnitude diagram of M$\,$67 in which the stars
detected in X-rays are shown with special symbols.
The size of the symbol is proportional to the logarithm of the
X-ray luminosity (in erg/s).
Circles indicate binaries with circular orbits, triangles binaries
with eccentric orbits, and squares binaries for which no orbit has been
determined. $+$ indicates an X-ray source for which no indication of binarity
has been found. From Belloni et al.\ (1998).
}
\end{figure}

\subsubsection{S$\,$1063 and S$\,$1113: stars below the subgiant branch}

The next two brightest X-ray sources in M$\,$67 are two stars which are
located below the subgiant branch in the colour magnitude diagram of
M$\,$67, number 1063 and 1113 in Sanders' list.
Both are binaries, one with a circular orbit, and one with an eccentric
orbit (Table~1).
The binary periods are too long, and the binaries are too far above the
main sequence, for these stars to be contact binaries.
Optical spectra show emission cores in the Ca$\,$II$\,$H and K lines
of both binaries (Van den Berg et al.\ 1998b), suggesting that the
X-ray emission could be due to magnetic activity. In addition there is
evidence for H$\,\alpha$ emission (Figure~8).

\begin{figure}
\psfig{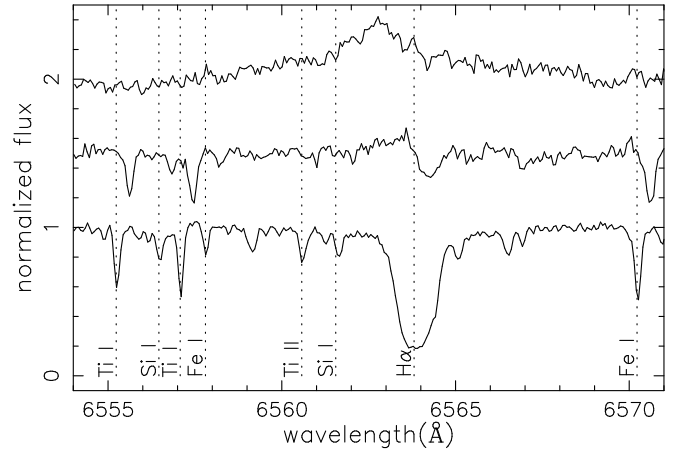}
\caption[]{H$\,\alpha$ line profiles for (from top to bottom)
the sub-subgiant binaries S$\,$1113, S$\,$1063 and for comparison
the single ordinary giant S$\,$1288 in M$\,$67. 
S$\,$1113 has the H$\,\alpha$ line in pure emission, whereas the
H$\,\alpha$ line of S$\,$1063 has been asymmetrically filled in with
emission.
Some line identifications are shown at the radial velocity of S\,1288.
Adapted from Van den Berg et al.\ (1998b).
}
\end{figure}

We do not understand the location of these stars below the subgiant
branch. In principle, mass transfer can cause a star to become
subluminous (because readjusting hydrodynamical equilibrium
for a mass-losing star drains its stellar luminosity), but
one expects very rapid circularization for a Roche-lobe-filling
star (Verbunt \&\ Phinney 1995).
Interestingly, the periastron distance in S$\,$1063 is such that
tidal forces can have brought a slightly evolved star into
corotation at periastron without (as yet) having circularized
the orbit (Van den Berg et al.\ 1998a); this could explain
the chromospheric emission of the spun-up star, but not why it is
underluminous.\nocite{bvm98b}
The X-ray flux of S$\,$1063 is variable, being $8.1\pm0.9$ counts/ksec 
in November 1991, and $4.7\pm0.6$ counts/ksec in April 1993 (Belloni et
al.\ 1998).

\subsubsection{S$\,$1072 and S$\,$1237: wide eccentric binaries}

Two X-ray sources are wide binaries, with orbital periods of about
1500 and 700$\,$d respectively, and with significant eccentricities.
Both binaries are in the yellow straggler region of the colour magnitude
diagram.
Their optical spectra do not show significant emission in the Ca$\,$II$\,$H and
K lines (Van den Berg et al.\ 1998b).
The eccentricity of the orbits indicate that no significant tidal interaction
is occurring or has occurred in these binaries.
We do not understand why these two binaries would emit X-rays.

\subsubsection{S$\,$1082: a blue straggler binary}

Star 1082 in Sanders' list is a blue straggler, the only one in this
cluster to be detected in X-rays.
From an ultraviolet excess, Landsman et al.\ (1998) conclude that 
it is a binary with a subdwarf O star.
Optical spectra taken by Van den Berg et al.\ (1998b) show variability
in the H$\,\alpha$ line, which may be related to the binarity.
The radial velocity of the star shows some variation
(Mathieu et al.\ 1986), but no orbital period
has been found so far; the orbital period of about 1 day suggested from
photometric data by Goranskii et al.\ (1992) in particular is not
present in the radial velocity data, contrary to expectation in
a short-period eclipsing binary.\nocite{lbn+98}\nocite{gkm+92}
Again, we do not know why this binary emits X-rays.

\subsection{NGC$\,$188}

Another old open cluster, NGC$\,$188, has also been observed with ROSAT,
and two sources have been detected in it (Belloni et al.\ 1998).
Their location in the colour magnitude diagram of NGC$\,$188 is shown
in Figure~9.
The brightest source, at 
$L_x({\rm 0.1-2.4\,keV})\simeq1.7\times10^{31}\,$erg/s,
is a rapidly rotating single giant, a star of the FK~Com type. 
Such stars possibly are the result of a merger of two stars into
a single, rapidly rotating star.
The other X-ray source, three times less luminous,
is identified with a star which is a short-period
velocity variable, as well as a rapid rotator.
The X-rays of both detected members of NGC$\,$188 can therefore be
explained in terms of magnetic activity of rapidly rotating
cool stars.

\begin{figure}
\psfig{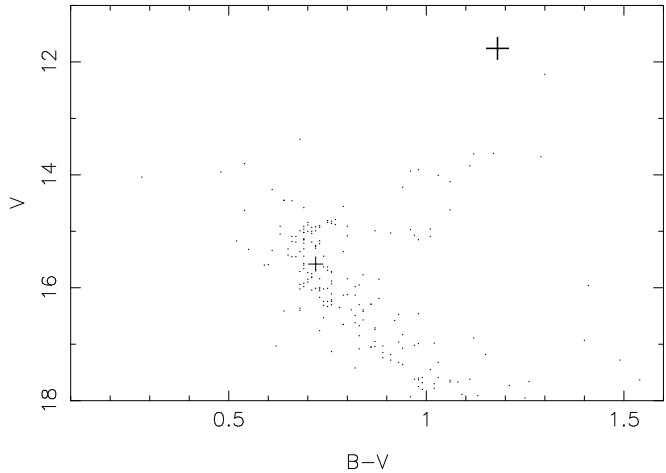}
\caption[]{Colour-magnitude diagram of NGC$\,$188 in which the stars
detected in X-rays are shown with $+$. From Belloni et al.\ (1998).
}
\end{figure}

\section{Summary, questions and outlook}

Whereas the nature of the bright X-ray sources in globular clusters
is clear -- at least 11 of the 12 known are accreting neutron stars
-- the nature of many of the less luminous sources is still unknown.
At least one dim source is a radio pulsar; one (in M\,5) is likely
a dwarf nova, and several others (in 47\,Tuc and in NGC\,6397) quite
possibly are cataclysmic variables.

The comparison of the old open clusters with the globular clusters
shows that the open clusters are surprisingly bright in X-rays.
The added X-ray luminosity of the M$\,$67 sources in the $0.5-2.5\,$keV
band is about $3\times 10^{31}\,$erg/s.
Note that this doesn't include the cataclysmic variable, which emits
only at energies $<0.5\,$keV.
If a typical globular cluster contains more than a thousand times as many
stars, with most binaries in the core, simple scaling from M$\,$67
would lead to a predicted X-ray luminosity for a globular cluster
well in excess of $10^{34}\,$erg/s, in clear contradiction to many
of the upper limits determined in the ROSAT All Sky Survey (see Figure~2).
Apparently, the X-ray sources of the types seen in M$\,$67 are 
relatively under-represented in the cores of globular clusters
(Verbunt 1996). \nocite{ver96d}
The same reasoning also shows that globular cluster cores do not contain
thousand FK~Com type stars with X-ray luminosities like that of the
one in NGC$\,$188.

Another remarkable fact uncovered by the analysis of M$\,$67 is that
none of the four brightest X-ray sources in this old open cluster
is of a known type of X-ray emitter.

This raises two questions for future research.
First, we would like to understand what causes the X-ray emission
in each of the four brightest X-ray sources in M$\,$67.
In particular it will be interesting to know whether the X-ray
emission of these stars is the consequence of more-or-less
ordinary binary evolution or whether the peculiar properties
of these binaries are the consequence of close encounters 
between binaries and/or single stars in the past history of M$\,$67.
N-body computations combined with stellar evolution of open clusters
may provide ideas (Aarseth \&\ Mardling 1996;
Portegies Zwart et al.\ 1997a,b). \nocite{am96}
\nocite{phv97}\nocite{phmv97}

Second, we would like to understand why X-ray sources like the 
brightest ones in M$\,$67 and NGC$\,$188 are not present in their
thousands in the cores of globular clusters.
In fact, even the less bright sources in the old open clusters,
like the RS CVn sources and the cataclysmic variables, cannot
be present in very large numbers in globular clusters.
The absence in globular clusters of wide binaries, and of binaries 
(like cataclysmic variables) that have evolved from wide binaries, 
can be understood from the efficiency with which wide binaries
are ionized in encounters with third stars (e.g. Davies 1997).
Theoretical formation and evolution scenarios of cataclysmic variables
in globular clusters  (Di Stefano \&\ Rappaport 1994) 
nonetheless predict much larger numbers than appear to be present
according to optical or ultraviolet observations
(Shara et al.\ 1996).\nocite{sr94}\nocite{sbg+96}
This indicates that our understanding of the formation (via tidal
capture) and of the evolution of cataclysmic variables, as
well as their phenomenology is far from complete.
The absence of close binaries, like RS~CVn systems, also is not readily
understood.
An alternative line of reasoning is that open clusters like M\,67
have already lost many of the stars originally in them, but 
retained a relatively larger fraction of the binaries.
Investigation into this line of reasoning again can be made with
N-body computations that include stellar evolution.

The near future will bring X-ray satellites that can study more 
X-ray sources in old open and globular clusters, like AXAF, 
or that can study the X-ray spectra of these sources, like XMM.
Together with theoretical and optical followup, these observations
will hopefully contribute answers to the questions about X-ray
sources in old clusters, that have been raised thanks to ROSAT.

\begin{acknowledgements}
I am grateful to the Leids Kerkhoven Bosscha Fonds for a travel grant
that enabled me to attend the meeting; and to Marten van Kerkwijk
for comments on the manuscript.
\end{acknowledgements}

\end{document}